\documentclass[aps,floatfix,prl,superscriptaddress, preprint]{revtex4-1}

\usepackage{tikz}
\usepackage{pgfplots}
\usepackage{gensymb}
\usepackage{graphicx}
\usepackage{caption}
\usepackage{upgreek}
\usepackage{textcomp}
\usepackage{here}
\pgfplotsset{compat=1.14}
\usepackage[colorinlistoftodos]{todonotes}

\begin{document}
\title{Micro-focusing of broadband high-order harmonic radiation by a double toroidal mirror}

\author{H\'el\`ene~Coudert-Alteirac}
\email{helene.coudert@fysik.lth.se}
\affiliation{Department of Physics, Lund University, P. O. Box 118, SE-22100 Lund, Sweden}
\author{Hugo~Dacasa}
\affiliation{Laboratoire d'Optique Appliquée, ENSTA ParisTech, École Polytechnique, CNRS - UMR7639, Chemin de la Huni\`ere, 91761 Palaiseau Cedex, France}
\author{Filippo~Campi}
\affiliation{Department of Physics, Lund University, P. O. Box 118, SE-22100 Lund, Sweden}
\author{Emma~Kueny}
\affiliation{Department of Physics, Lund University, P. O. Box 118, SE-22100 Lund, Sweden}
\author{Bal\'azs~Farkas}
\affiliation{ELI-ALPS, ELI-HU Non-Profit Ltd., Dugonics tér 13, Szeged 6720, Hungary}
\author{Fabian~Brunner}
\affiliation{Department of Physics, Lund University, P. O. Box 118, SE-22100 Lund, Sweden}
\author{Sylvain~Maclot}
\affiliation{Department of Physics, Lund University, P. O. Box 118, SE-22100 Lund, Sweden}
\author{Bastian~Manschwetus}
\affiliation{Department of Physics, Lund University, P. O. Box 118, SE-22100 Lund, Sweden}
\author{Hampus~Wikmark}
\affiliation{Department of Physics, Lund University, P. O. Box 118, SE-22100 Lund, Sweden}
\author{Jan~Lahl}
\affiliation{Department of Physics, Lund University, P. O. Box 118, SE-22100 Lund, Sweden}
\author{Linnea~Rading}
\affiliation{Department of Physics, Lund University, P. O. Box 118, SE-22100 Lund, Sweden}
\author{Jasper~Peschel}
\affiliation{Department of Physics, Lund University, P. O. Box 118, SE-22100 Lund, Sweden}
\author{Bal\'azs~Major}
\affiliation{ELI-ALPS, ELI-HU Non-Profit Ltd., Dugonics tér 13, Szeged 6720, Hungary}
\author{Katalin~Varj\'u}
\affiliation{ELI-ALPS, ELI-HU Non-Profit Ltd., Dugonics tér 13, Szeged 6720, Hungary}
\author{Guillaume~Dovillaire}
\affiliation{Imagine Optic, 18 Rue Charles de Gaulle, 91400 Orsay, France}
\author{Philippe~Zeitoun}
\affiliation{Laboratoire d'Optique Appliquée, ENSTA ParisTech, École Polytechnique, CNRS - UMR7639, Chemin de la Huni\`ere, 91761 Palaiseau Cedex, France}
\author{Per~Johnsson}
\affiliation{Department of Physics, Lund University, P. O. Box 118, SE-22100 Lund, Sweden}
\author{Anne~L'Huillier}
\affiliation{Department of Physics, Lund University, P. O. Box 118, SE-22100 Lund, Sweden}
\author{Piotr~Rudawski}
\affiliation{Department of Physics, Lund University, P. O. Box 118, SE-22100 Lund, Sweden}

\begin{abstract}
We present an optical system based on two toroidal mirrors in a Wolter configuration to focus broadband XUV radiation. 
Optimization of the focusing optics alignment is carried out with the aid of an XUV wavefront sensor. 
Back-propagation of the optimized wavefront to the focus yields a focal spot of 3.6$\times$4.0 $\upmu$m$^2$ full width at half maximum, which is consistent with ray-tracing simulations that predict a minimum size of 3.0$\times$3.2 $\upmu$m$^2$. This work is important for optimizing the intensity of focused high-order harmonics in order to reach the nonlinear interaction regime.
\end{abstract}
\maketitle

\section{Introduction}

High-order harmonic generation (HHG) \cite{McPherson1987,Ferray1988} in gases is one of the main sources of attosecond pulses in the extreme-ultraviolet (XUV) or soft X-ray regime. These pulses, generated through highly nonlinear interaction between laser pulses and atoms, are coherent and broadband, typically several tens of eV wide. If sufficiently intense, they can be used to investigate multiphoton phenomena in atoms and molecules to unravel ultrafast dynamics \cite{Tzallas2003,Tzallas2011,Nabekawa2005,Manschwetus2016}. 
These nonlinear experiments require high intensity, on the order of $10^{12}-10^{14}$ W/cm$^{2}$. 
While numerous studies report optimization of the pulse energy, up to the $\upmu$J level \cite{Hergott2002, Midorikawa2008, Rudawski2013}, as well as the generation of extremely short pulses of the order of 100 as or shorter \cite{Zhao2012,Li2017a}, few tackle the problem of focusing the HHG radiation \cite{Valentin2003,Morlens2006a}. The challenge is to focus an XUV beam to a small focal spot, while keeping the broad bandwidth of the radiation, the short pulse duration, and minimizing the loss in the pulse energy.

Refractive optics cannot be used for XUV radiation, since it is strongly absorbed when propagating in materials. 
Spherical mirrors, used at near-normal incidence, allow high-aperture beams to be focused to small focal spots. 
However, they are effectively designed only for narrow bandwidth radiation. 
Although larger bandwidths can be handled by aperiodic multilayer mirrors, their reflectivity is often very low \cite{Morlens2006a,Menesguen2010}. 
Diffractive optics, such as zone plates \cite{Wieland2003}, also have low transmission and are designed for monochromatic radiation. 
The best option to focus a broadband XUV beam with high reflectivity is therefore using curved mirrors at grazing incidence \cite{Poletto2013}. 
Ellipsoidal mirrors are in principle able to perfectly focus the radiation in the second focus of the ellipsoid, if the point source is located in the first focus.
An ellipsoidal mirror was used to focus harmonics between 24 and 38 nm (33 to 51 eV) with a full width at half maximum (FWHM) measured to be 2.5 $\upmu$m \cite{Suda2007}. 
However, these mirrors are challenging to manufacture as well as to align, since strong aberrations appear if the source is extended, like in the case of HHG, or if it moves slightly out of the first focus of the ellipsoid \cite{Bourassin-Bouchet2013}. 

Configurations with multiple toroidal mirrors have been used \cite{Poletto2013} to overcome the limitation of ellipsoidal mirrors. 
The concept is to mutually compensate the coma \cite{Underwood1998} with successive mirrors. 
A Z-shaped configuration with two toroidal mirrors was used to focus high-order harmonics between 30 and 70 nm (17 to 38 eV) to an 8 $\upmu$m FWHM spot \cite{Valentin2003, Frassetto2014}. 
In the 1950s, Wolter \cite{Wolter52} demonstrated that some combinations of two confocal conic sections, such as ellipsoid, paraboloid, and hyperboloid, can minimize aberrations. 
This idea is widely used in astronomy for X-ray telescopes \cite{Gaetz2000}, for focusing neutron beams \cite{Mildner2011,Khaykovich2011}, and for inertial confinement fusion imaging experiments \cite{Bennett2001,Ellis1990}. 

Focusing of XUV radiation requires not only good optics but also precise alignment and therefore accurate measurement techniques. 
The focus size can be determined by knife-edge \cite{LeDeroff1998}, point diffraction \cite{Lee2003}, slit diffraction \citep{Frumker2009}, lateral shearing interferometry \cite {Austin2011}, or direct imaging with a microscope \cite{Valentin2003}.
This direct technique requires the insertion of a component in the focus, which hinders simultaneous application experiments. 
Wavefront sensors \cite{Mercere2003}, on the other hand, can be located far from the focus and thus provide on-line diagnostics of the beam spatial characteristics. 
Shack-Hartmann sensors, based on microlens arrays, are routinely used in the visible or infrared regime, but their use in the XUV regime is more challenging \cite{LePape2002}. Therefore the Hartmann technique, based upon diffraction through small apertures, is often used \cite{Mercere2003}. 
This sensor allows to determine the wavefront aberrations after focusing, and thus provides convenient feedback for the alignment of the focusing optics. 
The focus location and spot shape can be estimated by performing back-propagation calculations.

In this work, we use two toroidal mirrors in a Wolter configuration to focus broadband high-order harmonic radiation between 20 and 50 eV. 
This design allows us to use a large demagnification of approximately 35 between the HHG source, located at 6 meters from the optics, and the image position at 170 mm from the mirror pair. 
We optimize the focusing of an XUV beam with an XUV wavefront sensor, using single-shot measurements, and back-propagate the measured intensity profile and wavefront to the focal plane which yields a spot of 3.6$\times$4.0 $\upmu$m$^2$ FWHM. 
These results are consistent with ray-tracing simulations that predict a minimum size of 3.0$\times$3.2 $\upmu$m$^2$ FWHM. 

In Section 2, we describe the experimental setup, including the XUV generation and focusing geometry. Wavefront measurements and focal spot optimization are presented in Section 3. We conclude in Section 4. 

\section{Experimental setup}

\subsection{Beamline}
\label{sec:beamline}

\begin{figure}[ht]
\includegraphics[width=\linewidth]{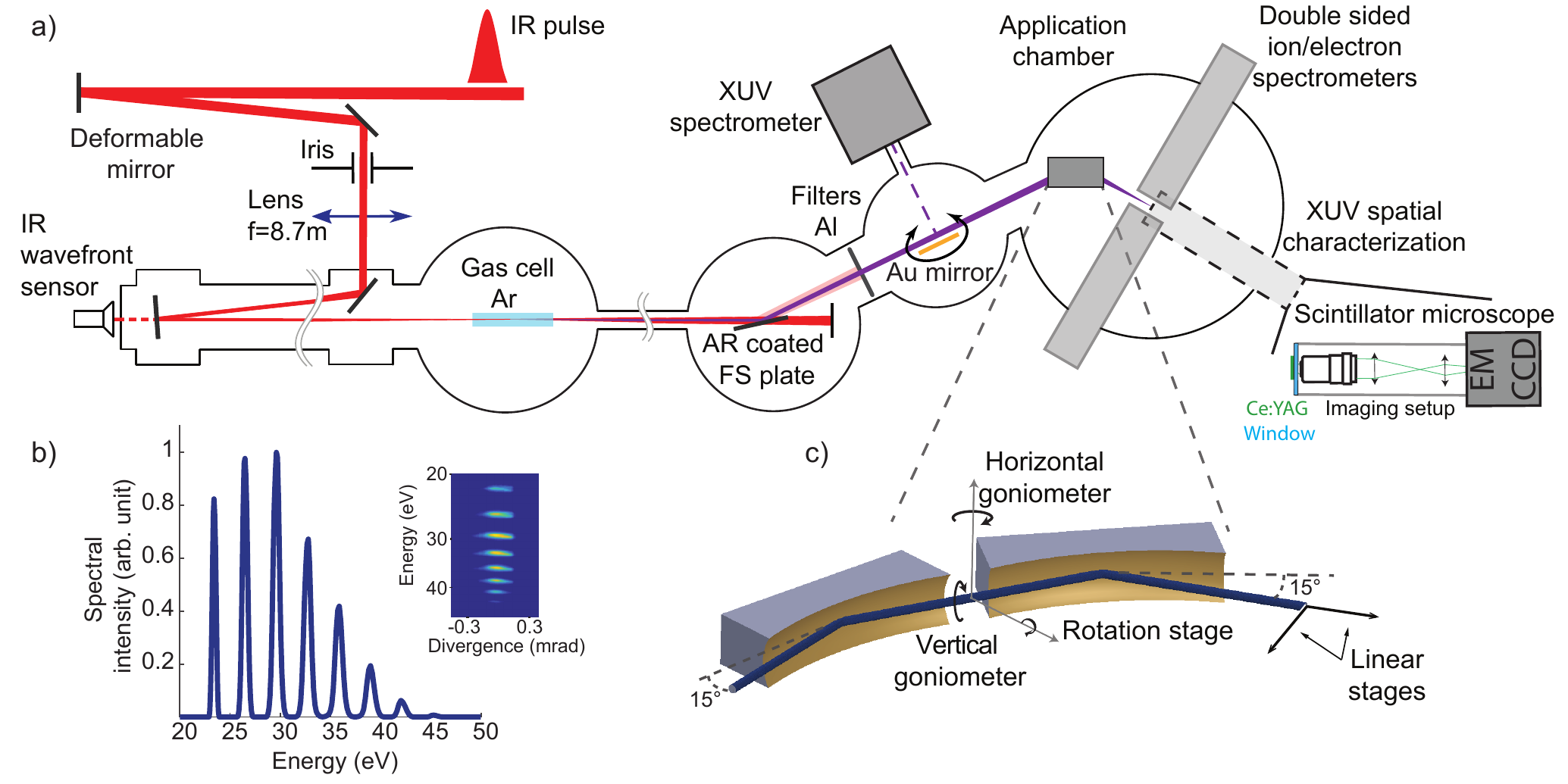}
\caption{a) Scheme of the intense XUV beamline, b) typical HHG spectrum and c) double toroidal mirror setup in a Wolter configuration. The black lines indicate the translation axes (straight solid arrows) and the pivoting axes for the revolutions (curved arrows).}
\label{Figure1}
\end{figure}

The intense XUV beamline \cite{Manschwetus2016} (Fig.~\ref{Figure1}a) is driven by a Ti:Sapphire chirped pulse amplification laser system delivering pulses at 10 Hz repetition rate with duration of 45 fs, energy up to 80 mJ, and a beam diameter of 24 mm (FWHM). 
The vertically polarized beam, apertured to a diameter of 22 mm, is focused by a 8.7 m focal length lens into a 60 mm long gas cell statically filled with argon. 
Prior to focusing, residual aberrations of the IR beam are compensated by a deformable mirror operating in closed loop with an IR wavefront sensor. The mirror also allows for adjustment of the focal position with respect to the cell in 1-mm steps. 
The aperture size, the gas pressure, the position of the cell, and the laser focus are optimized for the maximum of the XUV signal. 
The IR and XUV radiation are then co-propagating towards a fused silica (FS) plate placed at a 10$^{\circ}$ grazing angle. 
The plate is anti-reflection (AR) coated to transmit the IR while the XUV beam is reflected. 
The residual IR radiation is blocked by a 200 nm-thick aluminum filter. 
In the next chamber, a gold-coated mirror on a rotation stage is used to fold the XUV radiation towards an XUV flat-field spectrometer. A typical HHG spectrum is shown in Figure \ref{Figure1}b. It is composed of eight harmonic orders, from 15 to 31, with photon energies from 23 to 42 eV. Their spatial profiles in the far field are plotted in the inset. The divergence is estimated to 0.33 mrad (FWHM) for harmonic 19.
The transmission of the beamline from the generation cell to the entrance of the application chamber is estimated to be 22$\%$, for XUV photons with energies from 20 to 50 eV. This value is calculated using the measured transmission of the Luxel aluminum filter (34$\%$) and the calculated reflectivity of the AR-coated fused silica plate (the top layer of the coating is SiO$_2$ with reflectivity of 65$\%$). 
During nonlinear experiments, the XUV beam is focused into a target gas jet, and a double-sided ion/electron spectrometer is placed around the focus as shown in Figure \ref{Figure1}.

\subsection{Micro-focusing}

With the intent of maintaining the broad spectrum and high throughput, the XUV radiation is focused in the experimental region by means of two toroidal mirrors in grazing incidence configuration. 
The XUV beam is not recollimated prior to refocusing, but instead the object, i.e. the generation volume in the gas cell, is imaged into the interaction region in the application chamber.
The object arm is approximately 6 m, while the image arm is 170 mm, resulting in a demagnification of 35. The combination of the two toroidal mirrors is designed to achieve a total deflection of 60$^{\circ}$, arising from a 15$^{\circ}$ grazing incidence on each mirror.

Figure \ref{Figure1}c shows a schematic representation of the Wolter assembly. It is composed of two toroidal mirrors which are 30 mm long and 10 mm high. The center-to-center distance between the two mirrors is 30 mm. 
The mirrors are coated with gold because of its broadband reflectivity in the XUV range. 
In fact, between 20 and 50 eV, the gold layer yields a 61$\%$ reflectivity for s-polarized light and for a 15$^{\circ}$ grazing angle. 
The theoretical throughput of the optical setup is then 37$\%$. However, the roughness of the surface is not taken into account in this calculation and the real throughput is expected to be smaller.
The two mirrors were installed and pre-aligned by the manufacturer (Thales SESO). 
This assembly is mounted on a 5-axis stage, whose degrees of freedom are sketched in Figure \ref{Figure1}c. 
These axes are actuated by piezo-driven stick-and-slip positioners in open-loop configuration.

As mentioned in the introduction, ellipsoidal mirrors are in principle the ideal solution for imaging. However, the toroidal surface is a good local approximation to the ellipsoidal surface with the advantage that the two focal lengths, $f_T$ and $f_S$, corresponding to the tangential (the plane which includes both the optical axis and the normal to the mirror surface) and the sagittal (normal to the tangential plane and containing the optical axis) planes, respectively, are not coupled and can be set independently according to the following equations:
\begin{eqnarray}
\label{eq:toroidal}
f_T=\frac{R_T\sin{\theta}}{2} \,,\\
f_S=\frac{R_S}{2\sin{\theta}} \,,
\end{eqnarray} 
where $R$ denotes the radius of curvature and the subscripts $T$ and $S$ refer to the tangential and sagittal planes, respectively. $\theta$ is the grazing angle. Table \ref{Woltertable} shows the curvatures and focal lengths of the individual mirrors in the two planes, as well as the combined focal length.

\begin{center}
\begin{table}[H]
\centering
\caption{Radii of curvature (provided by the manufacturer) and focal lengths of the individual mirrors, in the two planes, and the equivalent focal length of the assembly.}
\label{Woltertable}
\begin{tabular}{c|c|c|c|c|}
\cline{2-5}
                                    & \multicolumn{2}{c|}{Tangential} & \multicolumn{2}{c|}{Sagittal}   \\ \cline{2-5} 
                                    & $R_T$ {[}mm{]}  & $f_T$ {[}mm{]} & $R_S$ {[}mm{]} & $f_S$ {[}mm{]} \\ \hline
\multicolumn{1}{|c|}{First mirror}  & 2050 & 265.3 & 137.2 & 265.0 \\  
\hline
\multicolumn{1}{|c|}{Second mirror} & 4213 & 545.2 & 281.8 & 544.4 \\
\hline
\multicolumn{1}{|c|}{Combined}       & \multicolumn{4}{c|}{Focal length = 164.2 mm}         \\ \hline
\end{tabular}
\end{table}
\end{center}
The curvatures of each of the two mirrors are adjusted so that the resulting focal lengths in both planes are almost the same, to prevent astigmatism. The equivalent focal length of the assembly is about 164 mm (from the center of the second mirror), calculated in the paraxial approximation. This implies that our source, located 6 m before the mirrors, is imaged at 170 mm from the center of the second mirror.

\subsection{Direct focus measurements}

In order to directly record an image of the focal spot, we used a Ce-doped yttrium aluminum garnet (Ce:YAG) scintillation crystal \cite{Moszynski1994, Valentin2003, Frassetto2014}, which converts the XUV radiation into visible light with 550 nm central wavelength. 
The resulting luminescence spot was imaged onto a cooled EMCCD (electron multiplying charge-coupled device) camera by using an optical system composed of a microscope objective (Mitutoyo M Plan Apo 100×) and two lenses for relay imaging (see inset to Figure \ref{Figure1}a).
The crystal was fixed to a thin fused silica window and placed under vacuum, while the imaging system with the camera was under atmospheric pressure. 
The field of view  and resolution were estimated to be 50 $\upmu$m and 2 $\upmu$m, respectively. 
A limiting factor for the measurement was the nonlinear dependence of the luminescence yield as a function of the XUV intensity \cite{Valentin2003,Jaegle1997} and, in particular, its saturation.  
To limit this phenomenon, we used a 2 $\upmu$m thick Al filter after the fused silica plate (see Figure \ref{Figure1}), which decreased the HHG energy by two orders of magnitude. Consequently, the luminescence yield was reduced, and the detection required a single-photon-sensitive camera.
Furthermore, it was crucial to record single-shot images of the focal spot, since accumulating several shots led to a seemingly larger spot size, because of the pointing instability of the incoming beam. 

These measurements confirm that the Wolter optics compensate efficiently for coma aberrations. Astigmatism could be observed in non-optimized focusing conditions. However, it was not possible to determine precisely the spot size, which was found to be systematically a factor two or three larger than the wavefront-based measurements presented in Section \ref{sec:wf measurements}. This probably indicates that this measurement method is still hampered by the nonlinearity of the crystal or other 
 technical limitations.

\section{Wavefront measurements and optimization of the focal spot size}
\label{sec:wf measurements}

\subsection{Hartmann wavefront sensor}

\begin{figure}[H]
\centering
\includegraphics[width=1\linewidth]{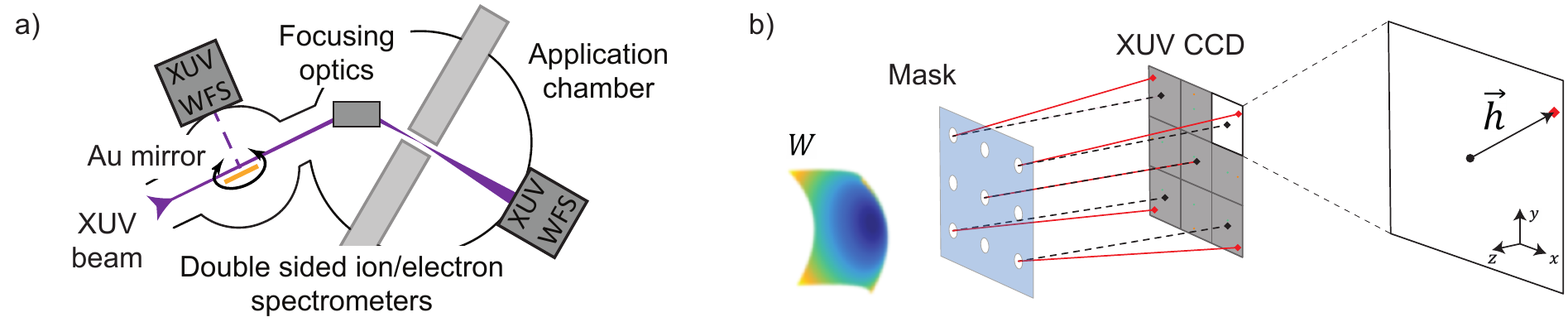}
\caption{a) Positions of the sensor in the beamline and b) principle of the Hartmann wavefront sensor.}
\label{fig_WFS}
\end{figure}

The wavefront sensor, presented schematically in Figure \ref{fig_WFS}a, is composed of a Hartmann mask and an XUV CCD camera \cite{Mercere2003}. 
Flat, non-tilted wavefronts result in individual diffraction spots being equidistantly distributed on the camera chip (see black dots in Figure \ref{fig_WFS}b), whereas in the case of distorted wavefronts, the positions deviate (red dots). These deviations are compared to a reference measurement with a known wavefront profile, here a spherical wavefront created from the diffraction of an HHG beam onto a pinhole. The local wavefront slopes in the mask plane, $(\partial W/\partial x,\partial W/\partial y)$ with $W$ denoting the wavefront, are proportional to these deviations ($\vec{h}$ in the figure), and are calculated for each hole. Subsequently, a wavefront reconstruction is performed to retrieve the wavefront over the whole aperture.

To measure XUV wavefronts after the focusing setup, we used a wavefront sensor (collaboration between Imagine Optics and the Laboratoire d'Optique Appliquée, LOA) placed 50~cm after the focus. This distance was chosen to match the beam size to the sensor aperture. The Hartmann mask is a 100~$\upmu$m thick nickel plate, which contains an array of 34$\times$34 holes. The holes are squares with a size of 110 $\upmu$m and a pitch of 387 $\upmu$m, and are tilted to prevent the diffraction patterns from consecutive holes to overlap. During previous calibration measurements performed at the LOA, the absolute accuracy of the sensor was estimated to be 0.6~nm RMS (root mean square), which corresponds to $\lambda / 70$ RMS for a wavelength of 42~nm. The sensor was mounted on a manual hexapod to ensure optimal alignment.
The analysis was performed for the 19th harmonic (42~nm~/~29.5~eV ), chosen as the weighted center of our spectrum. The validity of single-wavelength wavefront retrieval for broadband radiation in the visible regime was discussed with a Shack-Hartmann sensor \cite{Hauri2005} and was proven to be satisfying.

There are two main types of wavefront reconstruction from the measured slopes \cite{Southwell1980}.
The first one is called zonal reconstruction, which corresponds to the direct numerical integration of the local slopes. 
The other type is modal reconstruction, a decomposition of the wavefront in the basis of orthogonal polynomials, which correspond to a set of different aberrations. It thus allows the aberrations to be extracted independently, with the drawback of fitting them over a circular pupil, which is not always matching real beam profiles and leads to slightly underestimated aberrations \cite{Lakshminarayanan2011}.
The basis of orthogonal polynomials is provided by Zernike polynomials $Z_n^m(x,y)$, which are often the preferred choice to extract and study aberrations independently. 
The wavefront is then expanded as:
\begin{equation}
W(x,y)=\sum_{m,n} c_n^m Z_n^m(x,y),
\label{zernike_coeff}
\end{equation}
where $c_n^m$ are the  weights of the individual polynomials. The first coefficients in Eq.~\ref{zernike_coeff} are known as piston ($c_0^0$), x-tilt ($c_1^{-1}$), y-tilt ($c_1^{1}$), defocus ($c_2^0$), astigmatism at 0$\degree$ ($c_2^2$), astigmatism at 45$\degree$ ($c_2^{-2}$), coma at 0$\degree$ ($c_3^1$) and 90$\degree$ ($c_3^{-1}$). 
The coefficients can be determined by fitting the polynomials' derivatives to the measured local slopes.

For the analysis of the wavefronts presented in this work, when using the zonal method, the pupil was chosen to be the entire illuminated area, about 10$\times$10~mm$^2$. When using the modal method, a circular pupil of 2.3~mm radius was centered around the peak of the beam profile. The analysis was performed using the first 32 Zernike polynomials with a standard peak-to-valley (PV) normalization.
All wavefronts presented here are single-shot measurements. Contributions from defocus and tilts have been numerically removed.

\subsection{Alignment of the focusing optics}

In order to evaluate the sensitivity of the wavefront aberrations to the alignment of the focusing optics, we scanned the rotation stage and goniometers described in Figure \ref{Figure1}c while recording single-shot wavefronts. As there is no absolute characterization of the stage rotation angles, we express them in number of steps. We extracted the RMS of the wavefronts by zonal reconstruction, as well as the corresponding Strehl ratio \cite{Cheriaux2001}, which is defined as the ratio of the peak intensities of the focal distribution in the aberrated and unaberrated cases. A good approximation of the Strehl ratio $S$, used in the present work, is given by $S=\exp(-\sigma^2/\lambda^2)$, where $\sigma$ is the wavefront RMS \cite{Mahajan1983}. 

\begin{figure}[H]
\centering
\includegraphics[width=1\linewidth]{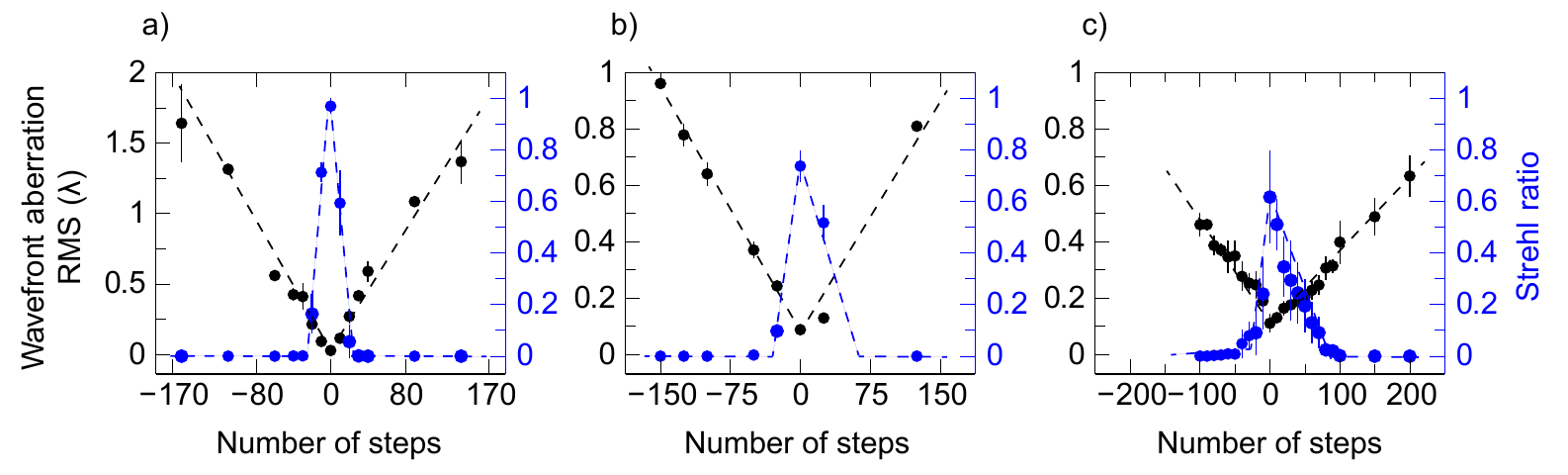}
\caption{Evolution of the XUV wavefront RMS as a function of the focusing optics relative angle (number of steps) along the different axes: a) rotation stage, b) horizontal goniometer and c) vertical goniometer. The RMS is plotted in black dots in units of lambda for $\lambda = 42$~nm (left axis), with error bars coming from several single-shot measurements. The corresponding Strehl ratios are plotted in purple dots (right axis). The qualitative trends are plotted in dashed lines.}
\label{actuators}
\end{figure}

The results are presented in Figure \ref{actuators}, where a, b and c correspond to scans of the rotation stage, the horizontal and vertical goniometers, respectively, with the other axes around the optimum. Due to shot-to-shot fluctuations, for each point, the average value of several (~5) single-shot measurements with corresponding RMS error bars is plotted.
A sharp V-shaped trend is visible for every graph. It means that for each actuator, it is possible to find a position minimizing the aberrations. The RMS minima are $0.028\pm0.015\lambda$ ($\simeq \lambda/36$) in a, $ 0.088\pm0.012 \lambda$ in b and $0.111\pm0.034 \lambda$  in c. The values for b and c are slightly higher because of  residual aberrations or misalignments.

\begin{figure}[H]
\centering
\includegraphics[width=1\linewidth]{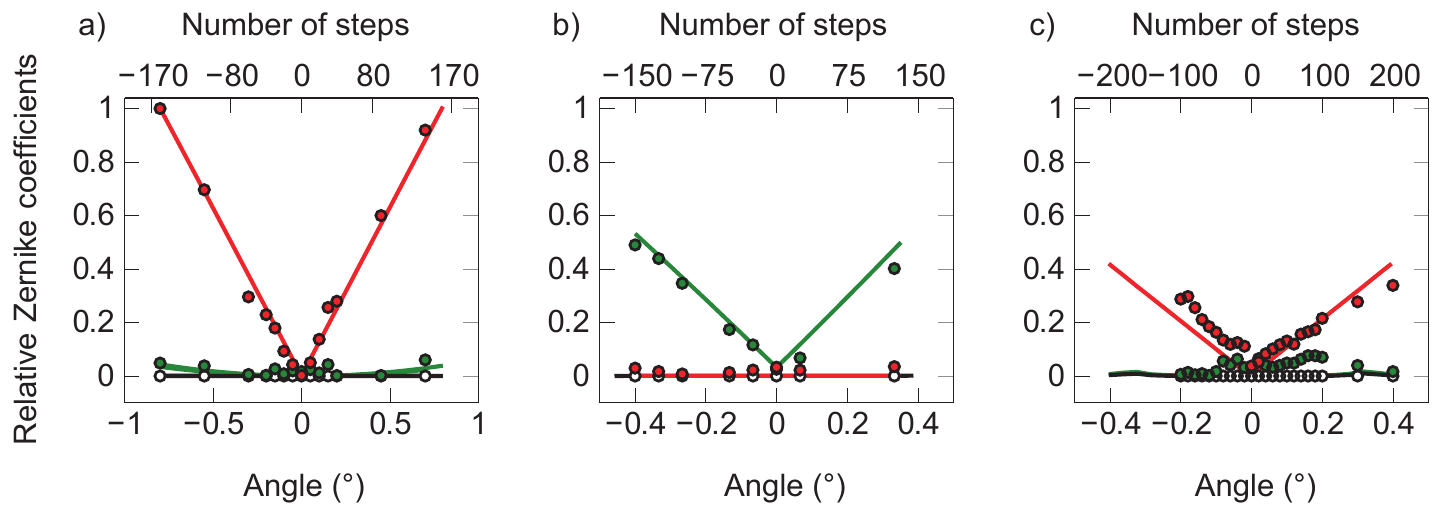}
\caption{Relative Zernike coefficients for the main aberrations as a function of the angle of the a) rotation stage, b) horizontal goniometer and c) vertical goniometer. The experimental data are represented as circled colored dots, plotted as a function of the number of steps (top axis). The ray-tracing simulation data is represented as solid lines, plotted as a function of the angle in degrees (bottom axis). Red corresponds to 45 $\degree$ astigmatism, green to $0\degree$ astigmatism, and white to the averaged coma.}
\label{zernike}
\end{figure}

In order to better understand which aberrations are accounting for most of the wavefront distortions, we used the modal technique to decompose the wavefronts in a weighted sum of Zernike polynomials, as explained above. Our analysis shows that astigmatism is the dominant aberration, accounting for most of the deviation from a perfect wavefront. Figure \ref{zernike} presents the Zernike coefficients for astigmatism (0$\degree$ and 45 $\degree$) and coma (average value of 0$\degree$ and 90$\degree$) as a function of the number of steps.
For all three cases, the coma is found to be negligible, which means that if the first mirror introduces coma, it is completely compensated by the second one, thus validating the double toroidal mirror Wolter configuration. The horizontal goniometer introduces mostly 0$\degree$ astigmatism while the rotation stage and the vertical goniometer introduce 45$\degree$ astigmatism. These different behaviors allow us to determine which axis is misaligned, and is thus very useful for fast optimization. For example, during the vertical goniometer scan shown in Figure \ref{zernike}c, there was residual 0$\degree$ astigmatism (green dots), meaning that the horizontal goniometer was not perfectly aligned in this case. Good alignment could be reached after a few iterations. 

In order to validate these results, we carried out simulations of the XUV beam propagation through the setup with the ray-tracing software FRED (from Photon Engineering), based upon Gaussian beamlet decomposition \cite{Harvey2015}. The wavelength was chosen to be 42~nm, corresponding to harmonic 19.  
The results are plotted in  Figure \ref{zernike} as solid lines, using the same colors as the experimental data for the different coefficients. Here, the relative angle is indicated in degrees. The ray-tracing simulations show the same behavior as the experimental data, with the sharp V-shape for the dependence of the main Zernike coefficient for a given axis. The simulations also confirm that each goniometer is affecting mainly a specific aberration. For the Zernike coefficients, simulations and experimental data have been normalized to 1 at -0.8 degrees (respectively -160 steps) of the rotation stage in Figure \ref{zernike}a. We also used the zonal wavefront reconstruction method for both experiment and simulation to extract the wavefront RMS. This allowed us to estimate the step size as a function of rotation angle for each axis.

\begin{figure}[H]
\centering
\includegraphics[width=1\linewidth]{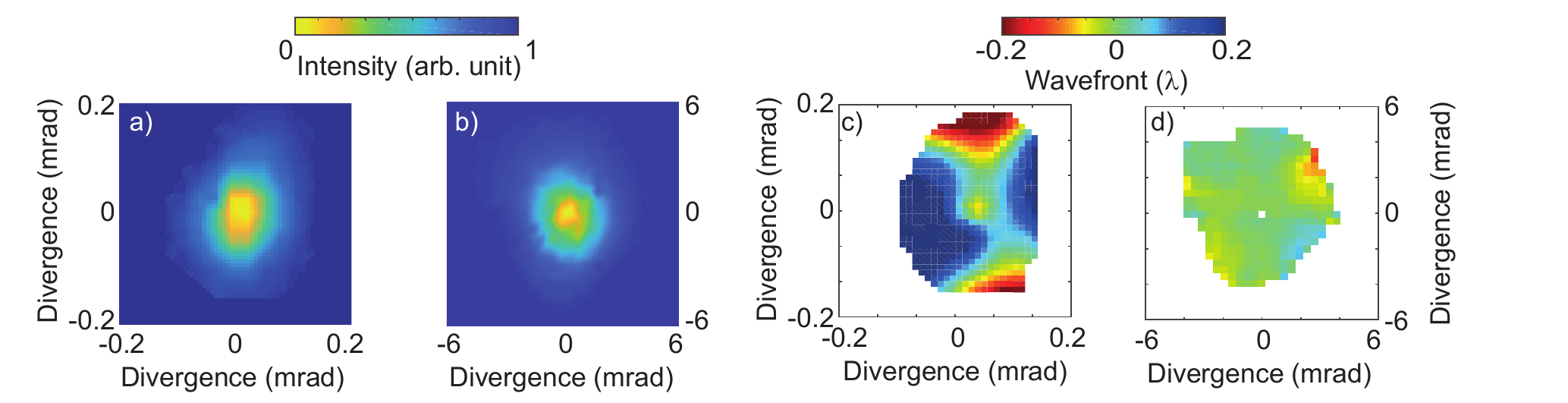}
\caption{Intensity and wavefront of the XUV beam before [a) and c)] and after [b) and d)] focusing.}
\label{before-after}
\end{figure}

Figure \ref{before-after} is a comparison between the beam before the focusing optics, intensity and wavefront in \ref{before-after}a and \ref{before-after}c respectively, and after optimization (\ref{before-after}b, \ref{before-after}d). The beam before focusing optics has some 0$\degree$ astigmatism, an RMS value of 0.11 $\lambda$ (=$\lambda$/9). After optimization of the focusing optics, we obtain an RMS value of $0.028\pm0.015 \lambda$ ($\lambda /36$).
It shows that the focusing optics can not only be aligned to introduce the least possible aberrations, but can also correct the pre-existing ones.

\subsection{Size of the focal spot}
\label{sec:back-prop}

To estimate the focal spot size, the measured wavefront and spatial intensity distribution are back-propagated to the focus by applying a Fourier transform. 
The wavefront $W(x,y)$ is written as the sum of the first three terms in the expansion (tilts and defocus, the piston is ignored) plus a residual wavefront $\tilde{W}(x,y)$.
The contributions of defocus and tilts to the wavefront are strong and usually dominate the higher-order aberrations. 
Their back-propagation requires a very fine sampling of the $xy$-plane over the pupil such that the wavefront never changes by more than a fraction of the wavelength from one sample point to the next. We thus need to treat the defocus and tilt contributions separately in order to optimize the numerical calculation.
The wavefront tilts represent the propagation direction of the beam and they can be added later to obtain the final result. The electric field $E(x,y,z)$, at the position of the Hartmann mask (z=0), is described by
\begin{eqnarray}
\label{eq:reformulation}
E(x,y,0)= U(x,y,0) \ e^{-ik c_2^0 Z_2^0(x,y)}\qquad \text{with}\qquad
U(x,y,0)= \sqrt{I(x,y,0)} \ e^{-ik\tilde{W}(x,y)}\ .
\end{eqnarray} 
$I(x,y,z=0)$ is the spatial intensity distribution measured at the position of the mask and $k=2\pi/\lambda$ the wavevector. Since the defocus component $c_2^0Z_2^0$ describes a quadratic phase, $E(x,y,0)$ represents the field after an ideal lens with focal length $f={r^2}/{(4c_2^0)}$, $r$ being the pupil radius, and $ U(x,y,0)$ being the field before the lens. The field in the focus is then given by the Fourier transform of $U(x,y,0)$
\begin{eqnarray}
\label{eq:backpropagation}
I(x',y',z_{foc}=f) = \vert E(x',y',z_{foc}) \vert ^2 \propto \vert \int U(x,y,0) e^{i2\pi(\nu _xx+\nu _y y)} dx dy \vert^2 .
\end{eqnarray}
Taking into account the tilts contribution to the focal spot position, the rescaled coordinates in the focal plane are $x'=\nu _x\lambda f+c_1^{-1}f/r$ and $y'=\nu _y\lambda f+c_1^{1}f/r$. The defocus component of the wavefront determines the focus position along the beam path, whereas the tilts determine its position in the $x'y'$-plane. 

\begin{figure}[H]
\centering
\includegraphics[width=0.8\linewidth]{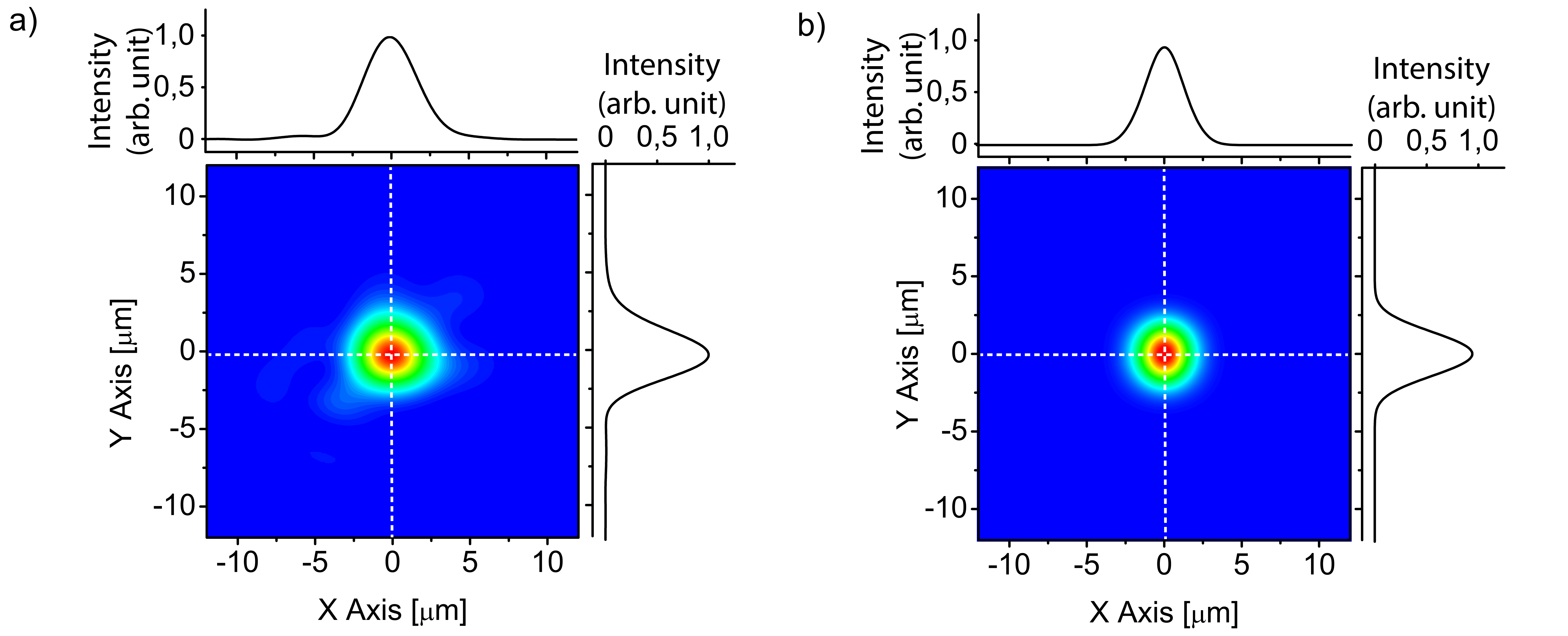}
\caption{Focal spot a) calculated from wavefront measurement by back-propagation, b) calculated from ray-tracing simulations.}
\label{fig_comparison}
\end{figure}

With this method, we back-propagated the least aberrated field (optimized alignment of the focusing setup), presented in Figure \ref{before-after}b and \ref{before-after}d, and obtained the focal spot plotted in Figure \ref{fig_comparison}a. The focus size is estimated to be 3.6$\times$4.0 $\upmu$m$^2$ (FWHM). 

We also performed ray-tracing simulations for aligned focusing optics, assuming a perfect gaussian beam, with the experimentally measured divergence. The focal spot is shown in Figure \ref{fig_comparison}b with its profiles. The FWHM beam size is 3.0$\times$3.2 $\upmu$m$^2$. 
The focal spot size obtained by back-propagation of the wavefront is in good agreement with the one obtained by the ray-tracing simulations. The 20\% difference in size can be explained by the difficulty of estimating precisely the divergence of the beam used for the ray-tracing simulations, and by distortions of the experimental intensity profile. 

The intensity of the focused attosecond pulse train is estimated to 7$\times 10^{12}$ W/cm$^2$, using the back-propagated spot size of 3.6$\times$4.0~$\upmu$m$^2$ and typical energies of 5 nJ, measured in the focal region. For the calculations we assume an equivalent pulse train duration to be 4.5 fs (train of 15 pulses with individual duration of $\approx 300$ as) \cite{Manschwetus2016}.

\section{Conclusion}

We have presented a micro-focusing setup for a broadband harmonic beam based on two toroidal mirrors in a Wolter configuration.
The setup is capable of focusing the beam to a spot size of 3.6$\times$4.0 $\upmu$m$^2$, which was optimized and characterized using an XUV wavefront sensor. The focus dimensions agree well with the expected value according to the ray-tracing simulations. The wavefront sensor was found to be an excellent tool to optimize the focusing optics alignment, providing quick and precise feedback. Pre-existing aberrations of the XUV beam could also be corrected.
The non-invasive property of wavefront measurements makes possible to control the focusing quality and thus guarantees the highest intensity on target during experiments.

\begin{acknowledgments}
The research was supported by the Swedish Research Council, the Swedish Foundation for Strategic Research, the European Research Council (grant 339253 PALP), LASERLAB-EUROPE (proposal LLC002276) and the European COST Actions MP1203 and CM1204. This project received founding from the European Union's Horizon 2020 research and innovation program under Marie Sk?odowska-Curie Grant Agreement no. 641789 MEDEA. The authors would like to thank Johan Axelsson and Anders Mikkelsen for lending high sensitivity cameras. The ELI-ALPS project (GINOP-2.3.6-15-2015-00001) is supported by the European Union and co-financed by the European Regional Development Fund. 
\end{acknowledgments}
%
%%%%%%%%%%%%%%%%%%%%%%%%%%%%%%%%%%%%%%%%%%
%\authorcontributions{
%H.C.-A., H.D., K.V., P.Z., P.J., A.LH, P.R. conceived and designed the experiments;
%B.F., H.C.-A., S.M., P.R. performed the spot characterization by the scintillator microscope;
%B.Man. designed the scintillator microscope;
%H.C.-A., H.D., P.R., F.C., E.K., S.M., H.W., J.L., L.R., J.P., B.Maj. performed the wavefront optimization experiments;
%F.C. and H.C.-A. performed the ray-tracing simulations;
%F.B. performed the back-propagation of the wavefronts and intensity recorded by the wavefront sensor;
%G.D. contributed to the analysis tools. 
%All authors participated in writing the manuscript.}
%
%%%%%%%%%%%%%%%%%%%%%%%%%%%%%%%%%%%%%%%%%%%
%\conflictsofinterest{The authors declare no conflict of interest. The funding sponsors had no role in the design of the study; in the collection, analysis, or interpretation of data; in the writing of the manuscript, and in the decision to publish the results.}

\bibliographystyle{apsrev4-1}
%\bibliography{Ref_lib}
\bibliography{microfocusing}

%%%%%%%%%%%%%%%%%%%%%%%%%%%%%%%%%%%%%%%%%%
\end{document}